\documentclass{PoS}

\usepackage{epsfig,multicol,multirow}
\usepackage{subfigure,latexsym}

\usepackage{graphics,amscd,amsfonts}
\usepackage{bm}
\usepackage{amsmath,amssymb}
\usepackage{comment}

\newcommand{\ba}{\begin{eqnarray}}
\newcommand{\ea}{\end{eqnarray}}
\newcommand{\beq}{\begin{equation}}
\newcommand{\eeq}{\end{equation}}

\newcommand{\eq} {equation}
\newcommand{\eqa} {eqnarray}
\newcommand{\NN} {\mbox {$\nonumber$}}

\allowdisplaybreaks

\title{Monte Carlo studies of 3d $\mathcal{N}=6$ superconformal Chern-Simons gauge
theory via localization method}

\ShortTitle{Monte Carlo studies of 3d $\mathcal{N}=6$ SCFT via localization method}

\FullConference{The 30th International Symposium on Lattice Field Theory\\
                 June 24 -- 29,  2012\\
                 Cairns, Australia}

\author{
\speaker{Masazumi Honda},$^{a,b}$\footnote{
This research was supported in part
by the National Science Foundation under
Grant No.~PHY11-25915.
M.~Honda and S.~Shiba are supported by Grant-in-Aid for JSPS fellows (No.~22-2764 and 23-7749).
The work of J.~Nishimura is supported in part by 
Grant-in-Aid for Scientific Research
(No.\ 20540286 and 23244057) from JSPS.
} Masanori Hanada,$^{b}$ 
Yoshinori Honma,$^{b,c}$ Jun Nishimura,$^{a,b}$ 
Shotaro Shiba$^{b}$ and Yutaka Yoshida$^{b}$
\vspace*{0.5cm} \\
\llap{$^a$}Department of Particle and Nuclear Physics,\\
Graduate University for Advanced Studies (SOKENDAI),\\
Tsukuba, Ibaraki 305-0801, Japan\\
\llap{$^b$}High Energy Accelerator Research Organization (KEK),\\
Tsukuba, Ibaraki 305-0801, Japan\\
\llap{$^c$}Harish-Chandra Research Institute,\\
Chhatnag Road, Jhusi, Allahabad 211019, India\\
\vspace*{0.5cm} \\
E-mail: \email{mhonda@post.kek.jp}, \email{hanada@post.kek.jp}, \email{yhonma@hri.res.in},\\
\email{jnishi@post.kek.jp}, \email{sshiba@post.kek.jp}, \email{yyoshida@post.kek.jp}}

\abstract{
We perform Monte Carlo study of the 3d $\mathcal{N}=6$ superconformal $U(N)\times U(N)$ Chern-Simons gauge theory (ABJM theory),
which is conjectured to be dual to M-theory or type IIA superstring theory on certain AdS backgrounds.
Our approach is based on a localization method, 
which reduces the problem to the simulation of a simple matrix model. 
This enables us to circumvent the difficulties in the original theory such as the sign problem and 
the SUSY breaking on a lattice. 
The new approach opens up the possibility of probing the quantum aspects of M-theory 
and testing the $AdS_4 /CFT_3$  duality at the quantum level. 
Here we calculate the free energy, and confirm the $N^{3/2}$ scaling 
in the M-theory limit predicted from the gravity side. 
We also find that our results nicely interpolate the analytical formulae proposed previously 
in the M-theory and type IIA regimes.\footnote{The 
simulation code is available upon request to mhonda@post.kek.jp.}
}

\begin{document}

\setcounter{footnote}{0}

\section{Introduction}
\vspace{-0.5em}
By now various regularization methods for supersymmetric gauge theories have been found, 
such as the lattice regularization (see e.g., ref.~\cite{Giedt:2009yd}), 
Fourier mode regularization \cite{Hanada:2007ti}, 
the large-$N$ reduction \cite{Eguchi:1982nm,Ishii:2008ib} and non-commutative geometry \cite{Hanada:2010kt}.
However, all these methods require 
a lot of computational costs due to the existence of the dynamical fermions. 
In this article 
we introduce a new simulation method for investigating 
a class of supersymmetric field theories 
via localization method \cite{Pestun:2007rz},
which reduces the evaluation of certain observables 
to calculations in simple matrix models. 
As a demonstration, we present numerical results \cite{Hanada:2012si} for 
the so-called ABJM theory \cite{Aharony:2008ug}, which is 
the 3d $\mathcal{N}=6$ superconformal $U(N)\times U(N)$ Chern-Simons gauge theory.

\vspace{-0.5em}
\section{Localization method for general 3d $\mathcal{N}=2$ supersymmetric field theory on $S^3$}
\vspace{-0.5em}
In this section
we explain the basic idea of the localization method \cite{Pestun:2007rz} and
write down the partition function
of a general 3d $\mathcal{N}=2$ supersymmetric field theory 
on $S^3$ in terms of a matrix model \cite{Kapustin:2009kz}. 
The ABJM theory belongs to this class of theories.
The localization method has been applied \cite{Pestun:2007rz} to 
4d ${\cal N}=4$ super Yang-Mills theory,
and some conjecture on the half-BPS Wilson loops
\footnote{This 
formula is also reproduced by a numerical simulation 
in the large-$N$ limit \cite{Honda:2011qk}.
} 
\cite{Erickson:2000af} has been confirmed. 
Those readers who are interested in just
understanding our numerical results may skip this section.

Let us consider the partition function of  a supersymmetric field theory,  
\vspace{-0.25em}
\begin{\eq}
Z=\int \mathcal{D} \Phi\ e^{-S[\Phi]} ,
\vspace{-0.25em}
\end{\eq}
where $\Phi$ represents the collection of the components fields. 
Let us suppose that the action is invariant under an off-shell supercharge $Q$, namely\footnote{
Here we assume the absence of the boundary term.
} $QS[\Phi ]=0$. 
Then, the closure of the SUSY algebra requires $Q^2 =\mathcal{L}_B$,
where $\mathcal{L}_B$ is the generator of a bosonic symmetry the theory has.
The first step of the localization method is to consider 
the deformation by a $Q$-exact term as
\vspace{-0.25em}
\begin{\eq}
Z(t) = \int \mathcal{D} \Phi\ e^{-S[\Phi] -t QV[\Phi ]},
\end{\eq}
where $V$ is any fermionic functional satisfying $\mathcal{L}_B V[\Phi ]=0$.
By taking the derivative with respect to $t$,
we obtain
\vspace{-0.25em}
\begin{eqnarray}
\frac{d Z(t)}{dt} 
= -\int \mathcal{D} \Phi\ (QV[\Phi ])e^{-S[\Phi] -t QV[\Phi ]}
&=& -\int \mathcal{D} \Phi\ Q\left( V[\Phi ]e^{-S[\Phi] -t QV[\Phi ]} \right)
\nonumber \\
&=& \int (Q\mathcal{D} \Phi )\  V[\Phi ]e^{-S[\Phi] -t QV[\Phi ]} .
\end{eqnarray}
If we assume the $Q$-invariance of the measure ($Q\mathcal{D} \Phi =0$), 
namely that $Q$ is non-anomalous,
then the deformed partition function $Z(t)$ should be independent of the parameter $t$. 
This implies that the original partition function $Z$ can be written as
\vspace{-0.25em}
\begin{\eq}
Z =\lim_{t\rightarrow +0} Z(t) =Z(t)
= \lim_{t\rightarrow\infty} \int \mathcal{D} \Phi\ e^{-S[\Phi] -t QV[\Phi ]} .
\vspace{-0.25em}
\end{\eq}
In this limit, the saddle point approximation around the classical solution to $QV=0$ becomes exact. 
Hence we obtain  
\vspace{-0.25em}
\begin{\eq}
Z = \sum_{\Phi_0}  \exp(-S[\Phi_0]) Z_{1-\mathrm{loop}}(\Phi_0) , 
\vspace{-0.25em}
\end{\eq}
where $\Phi_0$ is the `localized' configuration determined by $(Q V)[\Phi_0 ]=0$.
The summation $\sum_{\Phi_0}$ over the saddle points
should be understood as an integration
if the saddle points are labeled by continuous parameters.
The one-loop determinant $Z_{1-\mathrm{loop}}$ around $\Phi_0$ is given by
\vspace{-0.25em}
\begin{\eq}
Z_{1-\mathrm{loop}} 
= \lim_{t\rightarrow\infty} \left. \int \mathcal{D} (\delta \Phi )\ e^{-t QV[\Phi ]} 
   \right|_{\Phi =\Phi_0 +\delta\Phi}.
\vspace{-0.25em}
\end{\eq}
We can also use this method to calculate $Q$-invariant operators
such as supersymmetric Wilson loops \cite{Kapustin:2009kz}.

Let us apply the localization method 
to a general 3d $\mathcal{N}=2$ supersymmetric gauge theory on $S^3$
which is a Yang-Mills Chern-Simons gauge theory
with arbitrary gauge group 
$G=G_1 \times \cdots \times G_r$ and Chern-Simons levels
coupled to arbitrary number of $\mathcal{N}=2$ chiral multiplets 
with arbitrary representations and R-charge assignment\footnote{
More generally, we can also include mass and FI terms \cite{Kapustin:2009kz}.
}.
The formula for the partition function is obtained as \cite{Kapustin:2009kz}
\begin{\eq}
Z
=\frac{1}{|W|}\int \frac{d^{ {\rm rank}G_1} \sigma^{(1)}}{(2\pi )^{ {\rm rank}G_1}}  \cdots
        \frac{d^{ {\rm rank}G_r} \sigma^{(r)}}{(2\pi )^{ {\rm rank}G_r}}\ 
   \prod_{a=1}^r  \Delta_{\rm Vec}^{G_a} (\sigma^{(a)}) 
   \prod_\alpha \Delta_{\rm Mat}^{\mathcal{R}_\alpha } (\sigma ;q_\alpha ) , 
\label{general}
\vspace{-0.25em}
\end{\eq}
where $|W|$ is the order of the Weyl group of $G$, and
$\sigma^{(a)}$ is the Cartan part of the adjoint scalar in the vectormultiplet 
with the gauge group $G_a$ at the localization point.
$\Delta_{\rm Vec}^{G_a } (\sigma^{(a)} )$ represents
the contribution from the vector multiplet with the gauge group $G_a$ given by\footnote{
Note that this formula is independent of the Yang-Mills gauge coupling.
This is because we can choose $Q\cdot V[\Phi ]$
as the action of $\mathcal{N}=2$ super Yang-Mills theory itself.
Then the deformation parameter $t$ is nothing but the gauge coupling.
}
\vspace{-0.5em}
\begin{\eq}
\Delta_{\rm Vec}^{G_a} (\sigma^{(a)} )
= \prod_{\alpha^{(a)} \in \Delta_+} \Bigl[ 2\sinh{\frac{\alpha^{(a)} \cdot \sigma^{(a)} }{2}} \Bigr]^2 \cdot
   \exp{\left[{\frac{ik_a}{4\pi} \sigma^{(a)} \cdot \sigma^{(a)}} \right]},
\vspace{-0.25em}
\end{\eq}
where $\alpha^{(a)}$ labels the positive roots of $G_a$.
$\Delta_{\rm Mat}^{\mathcal{R}_\alpha } (\sigma ;q_\alpha )$ is
the contribution from the chiral multiplet with the representation $\mathcal{R}_\alpha$ and
R-charge $q_\alpha$ ( $q_\alpha =1/2$ in the canonical assignment) :
\vspace{-0.25em}
\begin{\eqa}
\label{DMat}
\Delta_{\rm Mat}^{\mathcal{R}_\alpha } (\sigma ;q_\alpha )
= \prod_{\rho_\alpha \in \mathcal{R}_\alpha} 
             f\left( i-iq_\alpha -\frac{\rho_\alpha \cdot \sigma}{2\pi} \right) ,
\vspace{-0.25em}
\end{\eqa}
where $\rho_\alpha$ is the weight vector of $\mathcal{R}_\alpha$ and $f(z)$ is given by
\vspace{-0.25em}
\begin{\eq}
f(z)
=\exp{\left[ -iz\log{(1-e^{2\pi z})}  
   -\frac{i}{2} \left( -\pi z^2 +\frac{1}{\pi}{\rm Li}_2 (e^{2\pi z}) \right) +\frac{i\pi}{12}
        \right]} .
\vspace{-0.25em}
\end{\eq}
As a special case of a pair of chiral multiplets 
with the representation $\mathcal{R}$ and $\bar{\mathcal{R}}$ 
in the canonical R-charge assignment,
which corresponds to the $\mathcal{N}=4$ hypermultiplet,
the formula (\ref{DMat}) reduces to the following simple form
\vspace{-0.25em}
\begin{\eqa}
\Delta_{\rm Mat}^{\mathcal{R} } (\sigma ; 1/2 )
\Delta_{\rm Mat}^{\bar{\mathcal{R}} } (\sigma ; 1/2 )
= \prod_{\rho \in \mathcal{R}} \frac{1}{2\cosh{\frac{\rho\cdot\sigma}{2}} } .
\vspace{-0.25em}
\end{\eqa}

\vspace{-0.5em}
\section{Numerical methods for the ABJM theory at
arbitrary $N$ and $k$}
\label{sec:methods}
\vspace{-0.5em}
Now let us consider the ABJM theory, which is 
the 3d $\mathcal{N}=6$ superconformal $U(N)\times U(N)$ Chern-Simons gauge theory \cite{Aharony:2008ug}. 
The Chern-Simons levels 
(the analogue of the gauge coupling constants) corresponding 
to two gauge groups are quantized to be integers, $k$ and $-k$. 
This theory is conjectured to be dual to M-theory 
on $AdS_4\times S^7/{\mathbb Z}_k$ for $k\ll N^{1/5}$,
and to type IIA superstring on $AdS_4\times {\mathbb C}P^3$ at $k\ll N\ll k^5$. 
The planar large-$N$ limit is defined as the large-$N$ limit 
with the 't Hooft coupling constant $\lambda=N/k$ kept fixed. 

The Monte Carlo study of the ABJM theory by usual lattice approach
seems quite difficult for the following three reasons. 
Firstly, the construction of the Chern-Simons term on the lattice
is not straightforward, although there is a proposal \cite{Bietenholz:2000ca}.
Secondly, the Chern-Simons term is purely imaginary in the Euclidean
formulation, which causes the sign problem in the importance sampling.
Thirdly, the lattice discretization necessarily breaks supersymmetry,
and one needs to restore it in the continuum limit by fine-tuning parameters\footnote{
This might be overcome by a non-lattice regularization of the ABJM theory \cite{Hanada:2009hd}
based on the large-$N$ reduction on $S^3$ \cite{Ishii:2008ib},
which is shown to be useful in studying the planar limit of the 4d $\mathcal{N}=4$
super Yang-Mills theory \cite{Honda:2011qk}.
}.

In order to circumvent these problems, we apply the Monte Carlo method 
to a matrix model obtained via the localization.
According to the general formula (\ref{general}),
the partition function of the ABJM theory on $S^3$ is given by
\vspace{-1.25em}
\begin{eqnarray}
\label{PF2}
Z(N,k)
&=&
\frac{1}{(N!)^2}\int\frac{d^N\mu}{(2\pi)^N}\frac{d^N\nu}{(2\pi)^N} 
\frac{\prod_{i<j}\Bigl[ 2\sinh\frac{\mu_i-\mu_j}{2}\Bigr]^2 
\Bigl[ 2\sinh\frac{\nu_i-\nu_j}{2}\Bigr]^2}
{\prod_{i,j} \Bigl[2\cosh\frac{\mu_i-\nu_i}{2}\Bigr]^2} 
e^{\frac{ik}{4\pi}\sum_{i=1}^N (\mu_i^2-\nu_i^2) } , 
\vspace{-0.25em}
\end{eqnarray}
which is commonly referred to as the ABJM matrix model.
From the partition function, we define the free energy as 
\vspace{-1.0em}
\ba
F(N,k)\,=\, \log Z(N,k) \,.
\label{defF}
\vspace{-1.0em}
\ea
Thus the ABJM free energy is given
just by a $2N$-dimensional integral. 
Note that the ABJM matrix model describes the continuum theory
without any regularization artifact.

The ABJM matrix model in the form \eqref{PF2} is not suitable 
for Monte Carlo simulation 
since the integrand is not real positive.
However, as we reviewed in Appendix~B of \cite{Hanada:2012si} 
in detail (See also the original work \cite{Kapustin:2010xq}), 
one can rewrite the ABJM matrix model as follows.
\vspace{-0.75em}
\begin{eqnarray}
&& Z(N,k)=C_{N,k}\,g(N,k) \,, \quad 
C_{N,k}=\frac{1}{(4\pi k)^N\,N!} \,, \NN \\
&& g(N,k)=
\int d^Nx\frac{\prod_{i<j}\tanh^2{\left( \frac{x_i-x_j}{2k}\right)}}
{\prod_i 2\cosh(x_i/2)} 
\equiv \int d^Nx\ e^{-S(N,k; x_1,\cdots ,x_N )} . 
\label{action_sign_free_1}
\vspace{-0.5em}
\end{eqnarray}
An important point here is that,
in this form (\ref{action_sign_free_1}), 
the integrand is real positive, and we can perform 
Monte Carlo simulation in a straightforward manner.

In order to calculate the partition function, 
we need to rewrite it in terms of expectation values of some quantities, 
which are directly calculable by Markov-chain Monte Carlo methods.
The basic idea
is to calculate the ratios of the partition functions
for different $N$ as expectation values\footnote{
We can also calculate the ratios of the partition functions
for different $k$ as expectation values \cite{Hanada:2012si}.
}.
Let us decompose $N$ into $N=N_1+N_2$ and consider the ratio
\vspace{-0.5em}
\begin{eqnarray}
\frac{g(N,k)}{g(N_1,k)g(N_2,k)} 
&=&
 \left\langle 
e^{S(N_1,k;x_1,\cdots,x_{N_1})+S(N_2,k;x_{N_1+1},\cdots,x_N)-S(N,k)}
\right\rangle_{N_1,N_2} \\
\label{reweight}
&=&  \left\langle 
\prod_{i=1}^{N_1} \prod_{J=N_1 +1}^N
              \tanh^2{\left( \frac{x_i -x_J}{2k} \right)}
\right\rangle_{N_1,N_2} \,,
\label{crucial-rel}
\vspace{-0.5em}
\end{eqnarray}
where the symbol $\langle \cdots \rangle_{N_1,N_2}$ 
represents the expectation value with respect to
the ``action'' given by
$S(N_1,k;x_1,\cdots,x_{N_1})+S(N_2,k;x_{N_1+1},\cdots,x_N)$.
In order to calculate the right-hand side of (\ref{reweight})
with good accuracy, 
it is necessary to take $N_2$ small enough to make sure that
the observable in (\ref{crucial-rel}) does not 
fluctuate violently during the simulation.  
In actual calculation we use $N_2 =1$.
Then, by calculating (\ref{reweight}) for $N_1 = 1,2,3,\cdots$
and by using the $N=1$ result $g(1,k)=\pi$,
we can obtain the free energy for $N=2,3,4,\cdots$ successively
with a fixed value of $k$.

\vspace{-0.5em}
\section{Results for the free energy}
\label{sec:FE}
\vspace{-0.5em}
\begin{figure}[t]
\begin{center} 
\includegraphics[width=7.4cm]{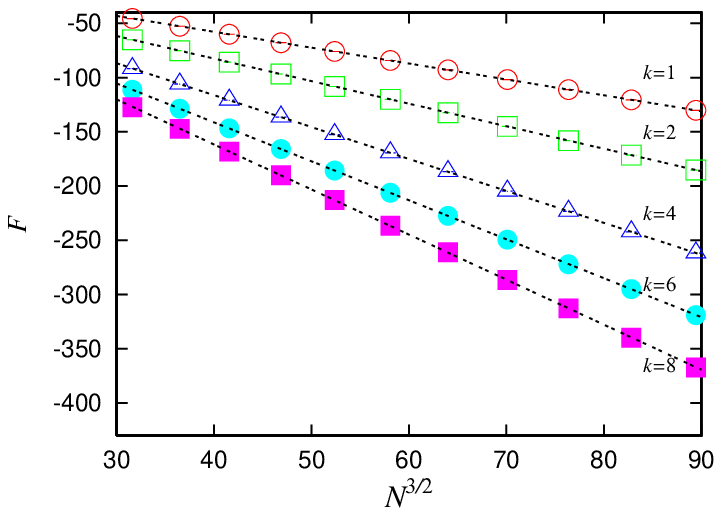} 
\includegraphics[width=7.4cm]{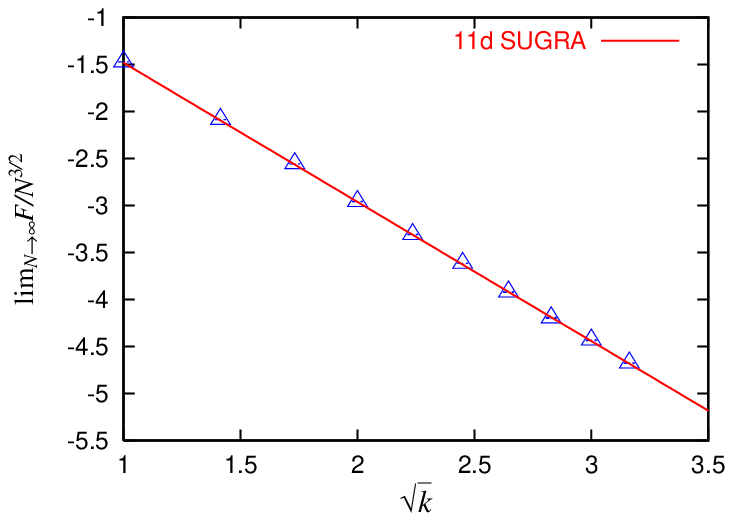} 
\end{center}
\caption{
(Left) The free energy is plotted against $N^{3/2}$ for
$k=1,2,4,6,8$. The data points can be fitted to straight lines,
which implies $F \sim N^{3/2}$ as $N$ increases.
(Right) The M-theory limit of the free energy $\lim_{N\rightarrow\infty}F/N^{3/2}$ 
is plotted against $\sqrt{k}$.
Our data are in good agreement with 
the result (5.1) predicted from the eleven-dimensional
supergravity, which is represented by the solid line.
}
\label{fig:M_extrapolation}
\end{figure}
We present our numerical result \cite{Hanada:2012si}
for the free energy of the ABJM theory.
First we consider the large-$N$ limit with fixed $k$, 
which is conjectured to correspond to the eleven dimensional supergravity 
on $AdS_4\times S^7/{\mathbb Z}_k$. 
In refs.~\cite{Herzog:2010hf,Marino:2011eh,Drukker:2010nc}, the free energy 
in the M-theory limit
($N\to\infty$ with $k $ \rm{fixed}) 
has been calculated by various analytic methods 
and confirmed the prediction
\vspace{-0.25em}
\begin{\eq}
F_{\rm SUGRA}=-\frac{\pi\sqrt{2k}}{3} N^{3/2} 
\label{SUGRA}
\vspace{-0.25em}
\end{\eq}
from the dual eleven-dimensional supergravity.
Figure~\ref{fig:M_extrapolation} (Left) shows that
the free energy $F$ grows in magnitude as $N^{3/2}$ with $N$.
Actually $F/N^{3/2}$ behaves as 
$F(N,k)/N^{3/2}=h_0 (k)+h_1 (k)/N$, 
which enables us to obtain the M-theory limit
$h_0 (k)=\lim_{N\to\infty}F(N,k)/N^{3/2}$ reliably.
In fig.~\ref{fig:M_extrapolation} (Right) we plot
$h_0 (k)$
against $\sqrt{k}$,
which confirms 
the prediction (\ref{SUGRA})
from the eleven-dimensional supergravity for $k=1,2,\cdots,10$ very precisely. 

\begin{figure}[t]
 \begin{center}
 \includegraphics[width=7.4cm]{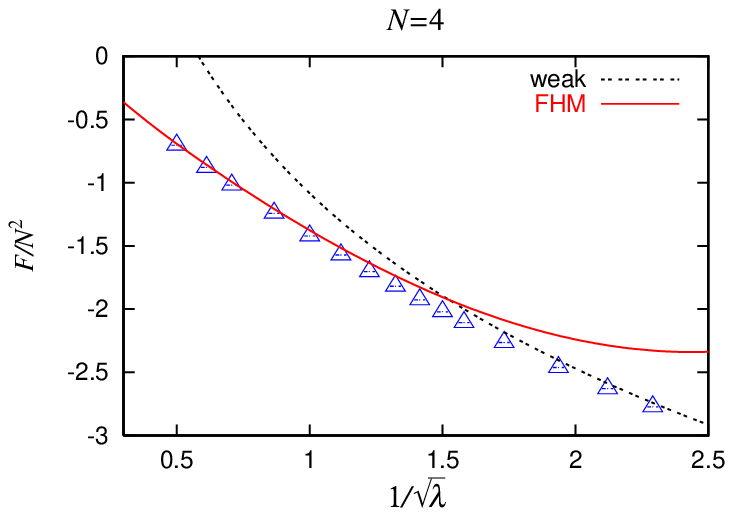}
\includegraphics[width=7.4cm]{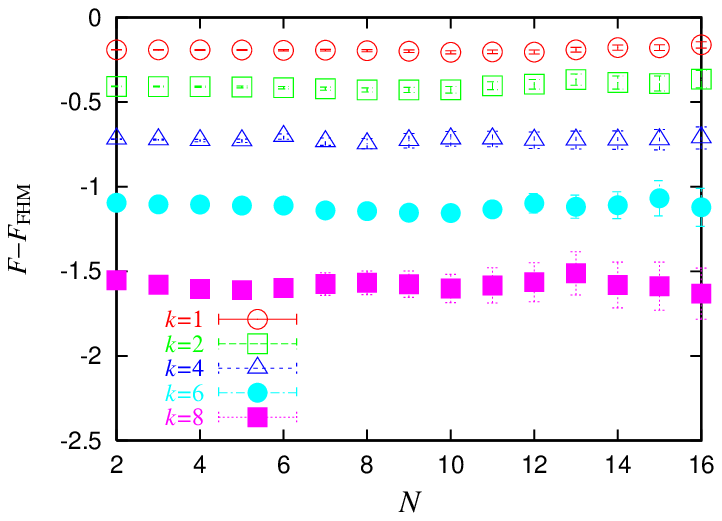}
 \end{center}
 \caption{(Left) The free energy of the ABJM theory for $N=4$ is
plotted against $1/\sqrt{\lambda}$. 
The solid line represents
the FHM result.
The dotted line represent the perturbative results
$F_{\rm weak} = -N^2\log\frac{2N}{\pi\lambda}-N\log 2\pi+2\log G_2(N+1)$
with the Barnes G-function $G_2(x)$. 
(Right) The difference $F-F_{\rm FHM}$ is plotted against $N$
for various values of $k$. 
It reveals non-negligible discrepancies for each $k$,
which are almost independent of $N$.
}
 \label{fig:free_energy_each_N}
 \end{figure}

Let us next study the finite-$N$ effects.
An important analytical result on finite $N$ effects is 
that the $1/N$ corrections around the planar limit
are resummed in a closed form \cite{Fuji:2011km,Marino:2011eh}
\vspace{-0.25em}
\begin{equation}
F_{\rm FHM}(N,\lambda ) \,=\, \log \left[
\frac{1}{\sqrt{2}}  \left( \frac{4\pi^2 N}{\lambda} \right)^{1/3} 
\mathrm{Ai}\left[\left( \frac{\pi N^2}{\sqrt{2} \lambda^2} \right)^{2/3}
\left( \lambda-\frac{1}{24} -\frac{\lambda^2}{3N^2} \right) \right]\right] \,, 
\label{FHM}
\vspace{-0.25em}
\end{equation}
where $\mathrm{Ai}(x)$ is the Airy function
and the type of correction ${\rm O}(e^{-2\pi \sqrt{\lambda}})$ is neglected.
In fig.~\ref{fig:free_energy_each_N} (Left) we plot our results for $N=4$ 
and compare them with the FHM result (\ref{FHM}).
We find 
that our result agrees reasonably well 
with the FHM result in the strong coupling regime.
To see more precisely,
we plot in fig.~\ref{fig:free_energy_each_N} (Right) 
the difference between our result and the FHM result against $N$ for various $k$. 
It turns out that
there are discrepancies which are almost independent of $N$.
This strongly suggests that the FHM result correctly incorporates 
the finite $N$ effects 
except for a term which depends only on $k$.
Note that this discrepancy cannot be explained by
the worldsheet instanton effect ${\rm O}(e^{-2\pi\sqrt{\lambda}})$,
which is neglected in FHM.
See ref.~\cite{Hanada:2012si} for 
a natural interpretation of this discrepancy from topological string theory.

\vspace{-0.5em}
\section{Summary and discussions}
\label{sec:con}
\vspace{-0.5em}

In this paper we have established a simple numerical method 
for studying
the ABJM theory on a three sphere for arbitrary rank $N$ 
and arbitrary Chern-Simons level $k$.
The crucial point is that we are able to rewrite the ABJM matrix
model, which is obtained after applying the localization technique,
in such a way that the integrand becomes positive definite.
By using this method, we have confirmed from first principles
that the free energy in the M-theory limit 
grows proportionally to $N^{3/2}$ as predicted from the
eleven-dimensional supergravity.
We have also found that 
the FHM formula with 
the additional terms 
describes the free energy of the ABJM theory
in the type IIA superstring and M-theory regimes. 
While we have focused on the free energy as the most fundamental
quantity in the ABJM theory, our method can be used
to calculate the expectation values of BPS operators.
For instance, it is possible to calculate the expectation value 
of the circular Wilson loop for various representations \cite{Wilson}. 

We hope that the results of this work are convincing enough
to show the power of the combination of the localization method and numerical simulation. 
We expect further numerical study of various localized matrix models will reveal exciting new aspects of supersymmetric gauge theories and quantum gravity.



\end{document}